\documentclass[fleqn,12pt,twoside]{article}
\usepackage[headings]{espcrc1}

\readRCS
$Id: espcrc1.tex,v 1.2 2004/02/24 11:22:11 spepping Exp $
\usepackage{graphicx}
\usepackage[figuresright]{rotating}

\newcommand{\AmS}{{\protect\the\textfont2
  A\kern-.1667em\lower.5ex\hbox{M}\kern-.125emS}}

\title{Strange and charged particle elliptic flow in Pb+Au collisions at 158~AGeV/c}

\author{J. Milo\v{s}evi\'c (for the CERES Collaboration) \\
        Physikalisches Institut der Universit\"{a}t Heidelberg, D 69120, Heidelberg, Germany}

\runtitle{Strange and charged particle elliptic flow in 158~AGeV/c Pb+Au collisions}
\runauthor{J. Milo\v{s}evi\'c}

\begin{document}

\maketitle

\begin{abstract}
We present $\Lambda$ and $\pi$ elliptic flow measurements from Pb+Au collisions
at the highest SPS energy. The data, collected by the CERES experiment which
covers $\eta=2.05,2.70$ with full azimuthal coverage and wide $p_{T}$
sensitivity up to $3.5$ GeV/c, can be used to test hydrodynamical models and
show sensitivity to the EoS. The value of $v_{2}$ as a function of centrality
and $p_{T}$ is presented. Values of $v_{2}$ observed by STAR at RHIC are
larger by about $1/3$. Our measurements are compared to results from other SPS
experiments and to hydrodynamical calculations. Huge statistics allows for a
precise measurement of the differential pion elliptic flow.
\end{abstract}

\section{INTRODUCTION}

Elliptic flow is described by the differential second Fourier coefficient of
the azimuthal momentum distribution $v_{2}({\cal D})=\langle \cos(2\phi)
\rangle_{\cal D}$ \cite{Oll,Bar}. The brackets denote averaging over many
particles and events, and ${\cal D}$ represents a phase-space window in the
$(p_{T},y)$ plane in which $v_{2}$ is calculated. The azimuthal angle
$\phi$ is measured with respect to the reaction plane defined by the impact
parameter vector $\bf b$ and the beam direction. For non-central collisions
($b \neq 0$), $v_{2}$ is an important observable due to its sensitivity to the
EoS, and through it to a possible phase transition to the QGP. The
$v_{2}(\Lambda)$ is important because $\Lambda$ is a baryon and in comparison
to other particle species could be used to check the mass ordering effect and
to compare it with hydrodynamical predictions which depend on EoS
used. Testing the differential flow measurements of different particle species
against different scaling scenarios may yield additional information about the
origin of flow.

\section{EXPERIMENT}

The CERES experiment consists of two radial Silicon Drift Detectors (SDD), two
Ring Imaging CHerenkov (RICH) detectors and a radial drift Time Projection
Chamber (TPC). Due to its full azimuthal coverage close to mid-rapidity the
CERES spectrometer is ideally suited for elliptic flow studies. The SDDs
located behind the target were used for the tracking and vertex
reconstruction. The purpose of the RICH detectors is electron identification
and they were not used in this analysis. The $v_{2}(p_{T})$ was studied in the
range $0.05<p_{T}<3.5$~GeV/c using the TPC which is operated inside a magnetic
field, providing a precise determination of the momentum. A more detailed
description of the CERES experiment can be found in \cite{Mar}. As data, we
used 3$\cdot 10^{7}$ Pb+Au events at 158~AGeV/c collected in the 2000 data
taking period. $91.2\%$ of events were triggered on $\sigma/\sigma_{geo} \le
7\%$, while only $8.3\%$ events with $\sigma/\sigma_{geo} \le 20\%$.

\section{METHOD OF $\Lambda$ RECONSTRUCTION}

The $\Lambda$ particles were reconstructed via the decay chanel $\Lambda
\rightarrow p+\pi^{-}$ with a $BR=63.9\%$ and $c\tau_{\circ}=7.89$~cm. As
candidates for $\Lambda$ daughters, only those TPC tracks were chosen which
have no match to an SDD track. Partial PID was performed using $dE/dx$
information from the TPC by applying a window around the momentum dependent
Bethe-Bloch value for pions and protons, respectively. On the pair level, a pt
dependent opening angle cut is applied, in addition to a cut in the
Armenteros-Podalanski variables ($q_{T} \le 0.125$~GeV/c and $0 \le \alpha \le
0.65$) to suppress $K_{S}^{0}$. \footnotemark

\footnotetext{With these cuts optimal values for $S/B\approx0.04$ and
  $S/\sqrt{B}\approx500$ were obtained ($S$ stands for the signal and $B$ for
  the background).}

\vspace*{-0.9cm}
\begin{figure}[h!]
  \begin{minipage}[c]{.40\textwidth}
    \centerline{\includegraphics[height=8.0cm]{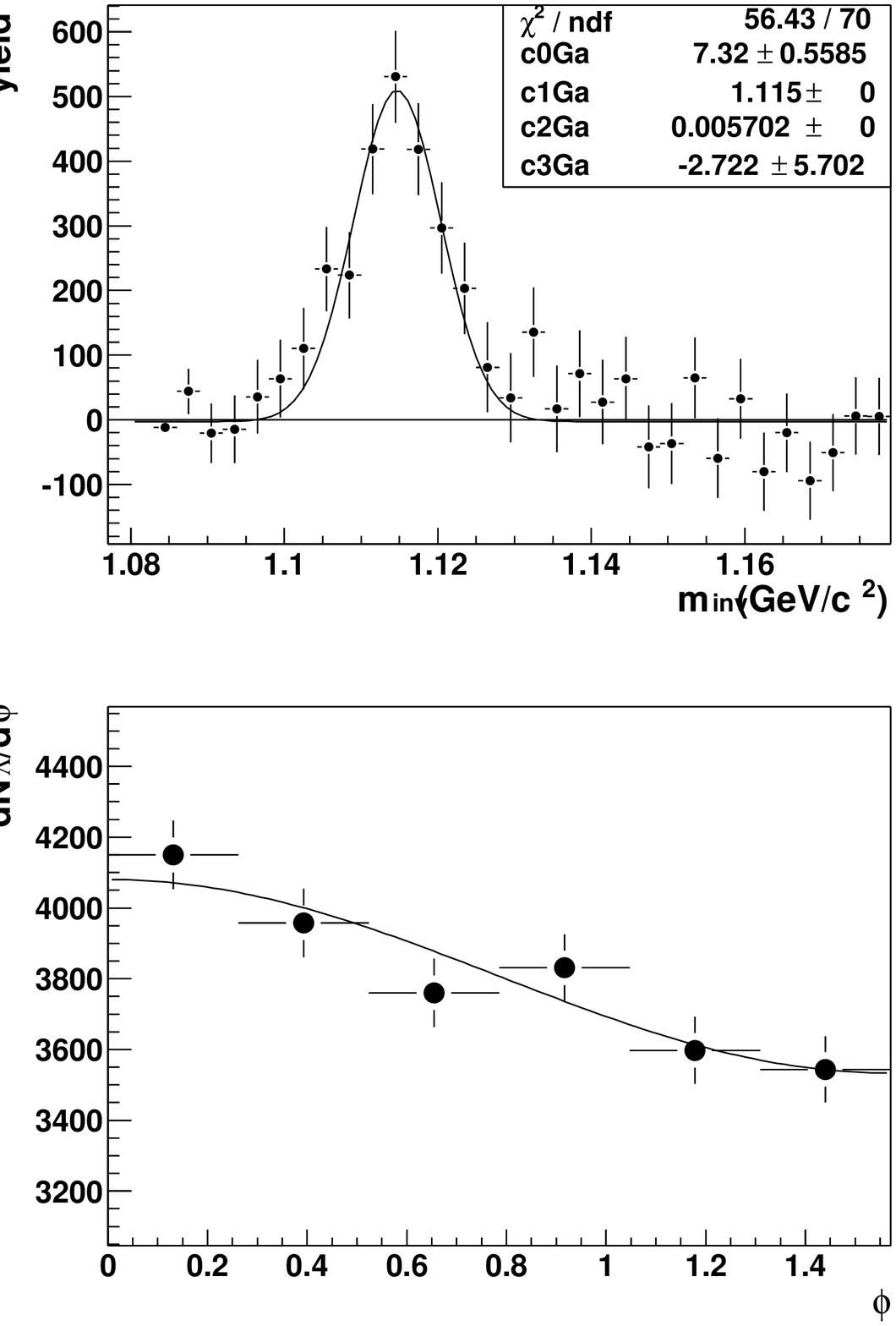}}
    \vspace*{-0.65cm}
    \caption{Top: $\Lambda$ reconstructed for $1.62\le y\le 1.69$,
    $0.675\le p_{T}\le 0.8$~GeV/c and $15^{\circ}\le \phi\le 30^{\circ}$.
    Bottom: Elliptic flow pattern reconstructed
    from the $\Lambda$ yield in $\phi$ bins for $p_{T} \approx
    2.7$~GeV/c.}
    \label{fig:example}
  \end{minipage}
  \hspace{\fill}
  \begin{minipage}[c]{.54\textwidth}
    \centerline{\includegraphics[height=8.0cm]{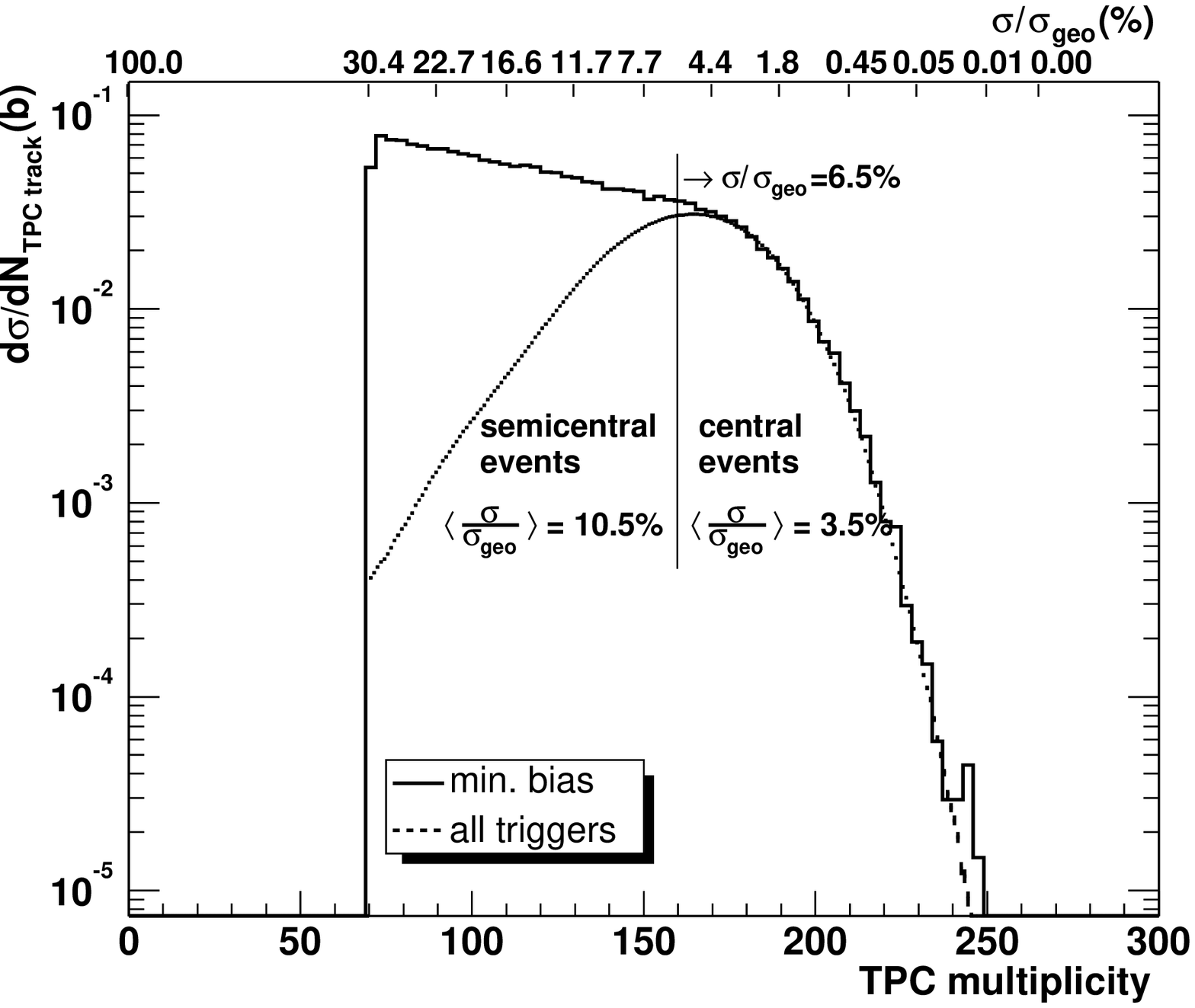}}
    \vspace*{-0.65cm}
    \caption{Determination of the centrality of events used in the
    analysis. The weighted mean centrality $\langle \frac{\sigma}{\sigma_{geo}}
    \rangle$ is calculated for 2 centrality bins. dotted: trigger data, solid:
    minimum bias sample.} 
    \label{fig:mult}
  \end{minipage}
\end{figure}

\vspace*{-0.6cm}
In order to remove the effect of autocorrelation, tracks which were chosen to
be candidates for daughter particles were not used for determination of the
reaction plane oriention. The combinatorial background was determined by
ten random rotations of positive daughter tracks around the beam axis and
constructing the invariant mass distribution.

An example of reconstructed $\Lambda$ in a given $y$-$p_{T}$-$\phi$ bin
is shown in Fig.~\ref{fig:example} (top). We used the area under the peak in
the invariant mass distribution to measure the yield of $\Lambda$s in the
given $y$-$p_{T}$-$\phi$ bin. Plotting the yield versus $\phi$ for different
$p_{T}$ and $y$ values one can construct $dN_{\Lambda}/d\phi$ distribution
(Fig.~\ref{fig:example}, bottom). Fitting these distributions with a function
$c[1+2v_{2}'\cos(2\phi)]$, it is possible to extract the observed elliptic flow
values $v_{2}'$ for different $p_{T}$ and $y$. The obtained $v_{2}'$
coefficients were corrected for the reaction plane resolution via
$v_{2}=v_{2}'/\sqrt{2\langle \cos[2(\Phi_{1}-\Phi_{2})]\rangle}$
\cite{Vol}. The resolution goes from 0.16 to 0.31.

\section{RESULTS}

\begin{figure}[h!]
\centerline{\includegraphics[width=15.0cm]{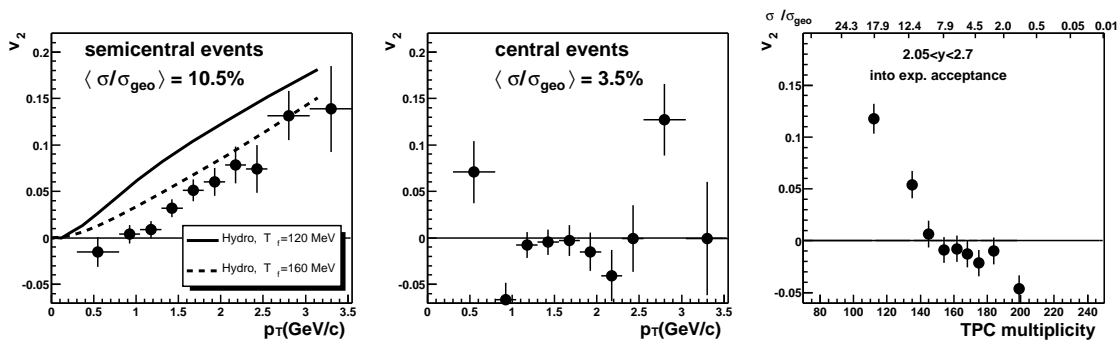}}
\vspace*{-0.65cm}
\caption{$p_{T}$ dependence of $v_{2}(\Lambda)$ for semicentral (left) and
  central (middle) collisions. Centrality dependence of the $\Lambda$
  elliptic flow (right).}
\label{fig:v2_QM}
\end{figure}

The elliptic flow analysis was performed for two centrality classes, defined
in Fig.~\ref{fig:mult}, and the resulting $p_{T}$ dependence of
$v_{2}(\Lambda)$ is shown in Fig.~\ref{fig:v2_QM}. One can see a clear
difference in the flow intensity between central and semicentral
\vspace*{-0.9cm}
\vspace*{-0.6cm}
events. The absolute systematic error $\Delta v_2$, estimated by varying the
applied cuts, is $+0.001 \atop -0.007$ for $p_{T} < 1.6$~GeV/c and $+0.00
\atop -0.02$ for $p_{T} > 1.6$~GeV/c which is small compared to the
statistical errors. The decrease of $v_{2}(\Lambda)$ with
$\sigma/\sigma_{geo}$ is displayed
\vspace*{-0.9cm}
\begin{figure}[h!]
  \begin{minipage}[c]{.48\textwidth}
    \centerline{\includegraphics[width=6.5cm]{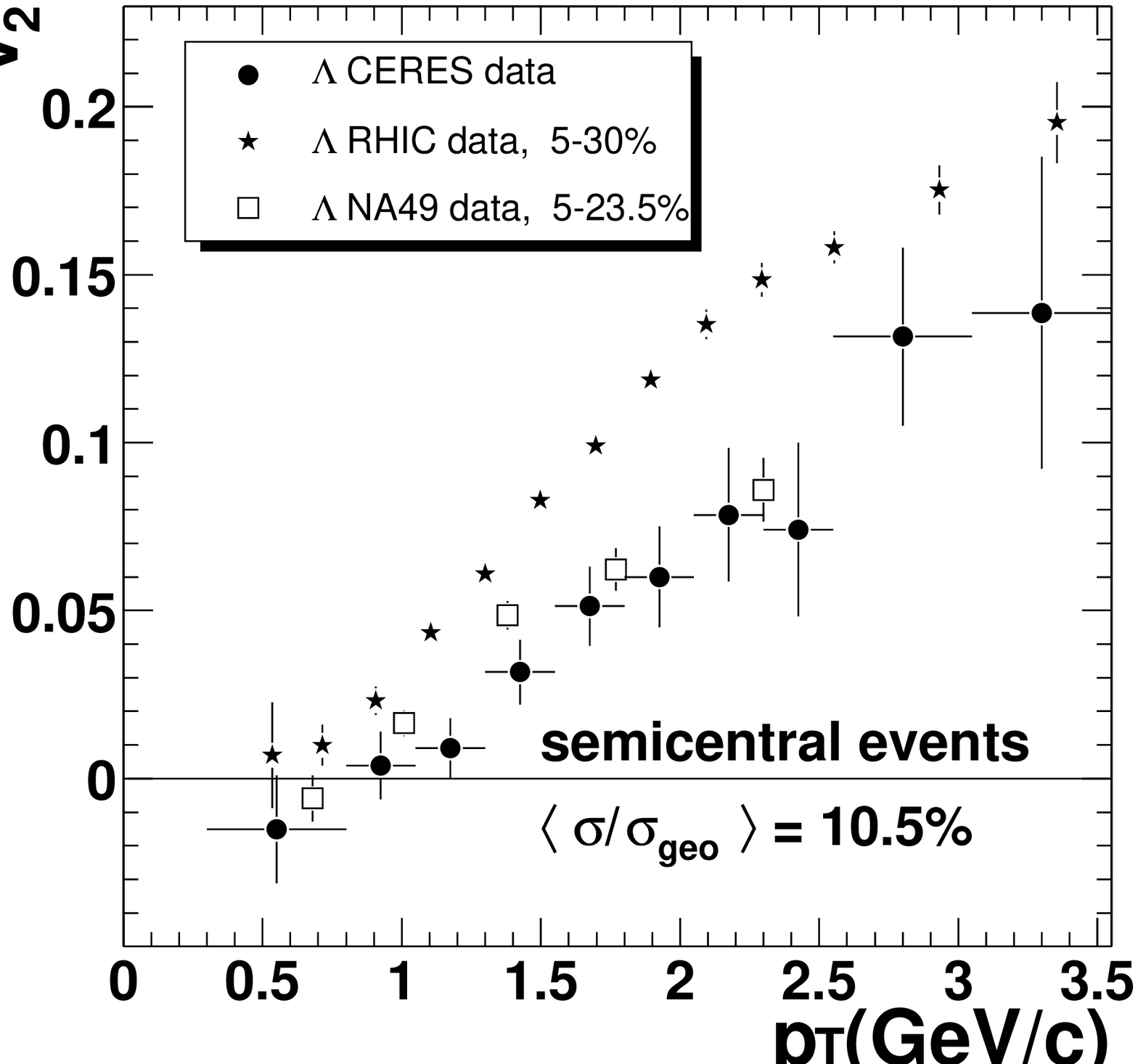}}
    \vspace*{-0.65cm}
    \caption{Comparison of $v_{2}(\Lambda)$ measured by CERES, STAR, and NA49.}
    \label{fig:SPS_RHIC_QM}
  \end{minipage}
  \hspace{\fill}
  \begin{minipage}[c]{.48\textwidth}
    \centerline{\includegraphics[width=6.5cm]{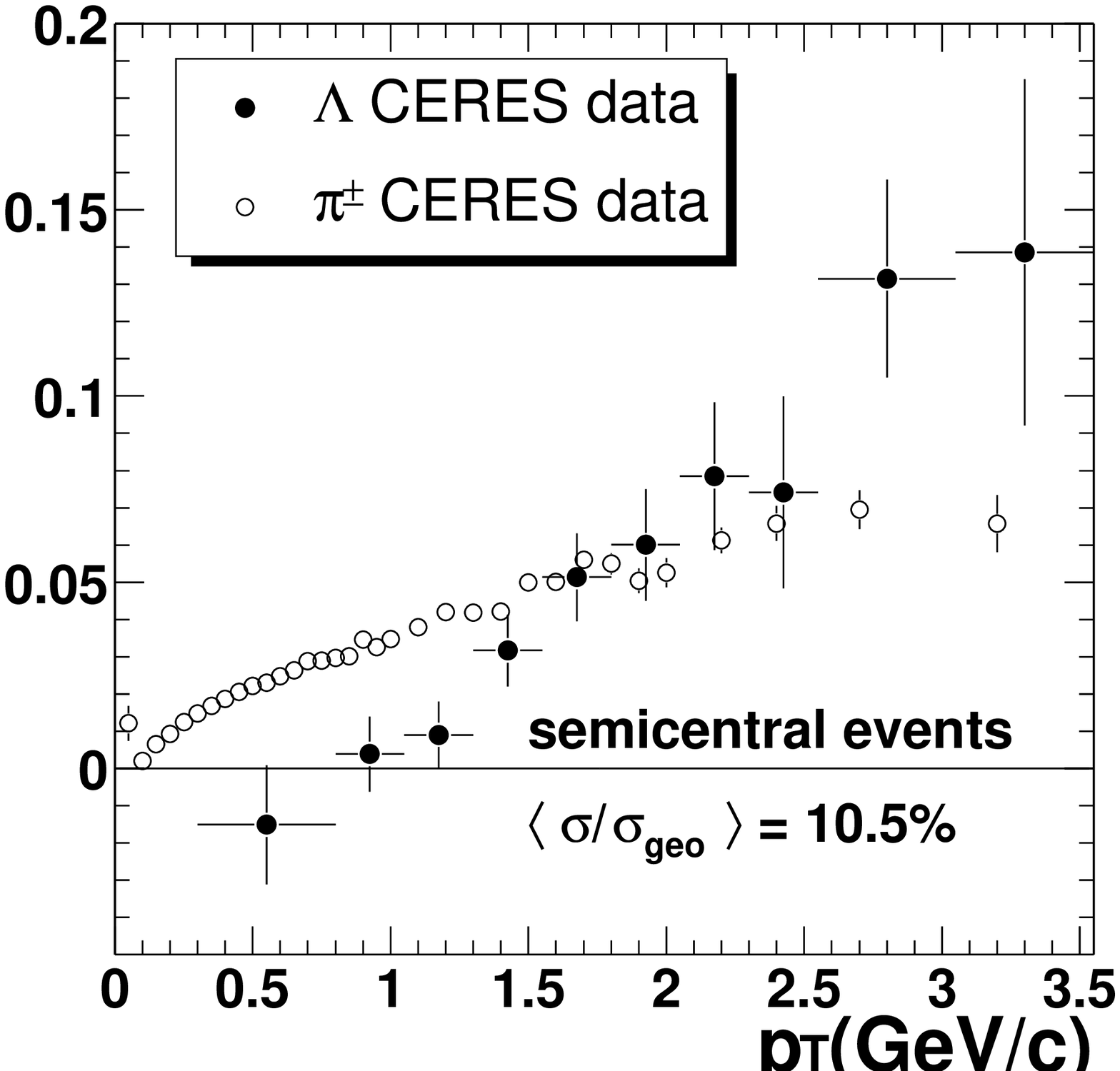}}
    \vspace*{-0.65cm}
    \caption{Comparison between $\Lambda$ and $\pi$ elliptic flow measured
    with CERES.}
    \label{fig:scaled_quark_1}
  \end{minipage}
\end{figure}
\vspace*{-0.6cm}
in Fig.~\ref{fig:v2_QM} (right). Fig.~\ref{fig:v2_QM} (left) also shows a
comparison between our results and hydrodynamical calculations \cite{Huo},
assuming a 1-st order phase transition at a critical temperature of
$165$~MeV. The model prediction with a higher freeze-out temperature of
$T_{f}=160$~MeV is close to the CERES data, while $T_{f}=120$~MeV overpredicts
the data. The same is observed comparing the pion flow from CERES to the same
hydrodynamical model \cite{Aga}.

A comparison of the CERES data to results from NA49 \cite{Stf} at the same
energy ($\sqrt{s_{NN}}=17$~GeV) and to STAR results \cite{Pos} at
$\sqrt{s_{NN}}=200$~GeV is shown in Fig.~\ref{fig:SPS_RHIC_QM}. The NA49 and
CERES data are in a very good agreement. The $v_{2}$ values measured at the
RHIC energy are $30-50\%$ higher. Partly, this is due to an effectively
higher centrality in CERES as compared to the STAR experiment.

The elliptic flow coefficients measured for $\Lambda$ and $\pi$ detected with
CERES are compared in Fig.~\ref{fig:scaled_quark_1}. At small $p_{T}$,
\vspace*{-0.9cm}
\begin{figure}[h!]
  \begin{minipage}[c]{.48\textwidth}
    \centerline{\includegraphics[width=6.5cm]{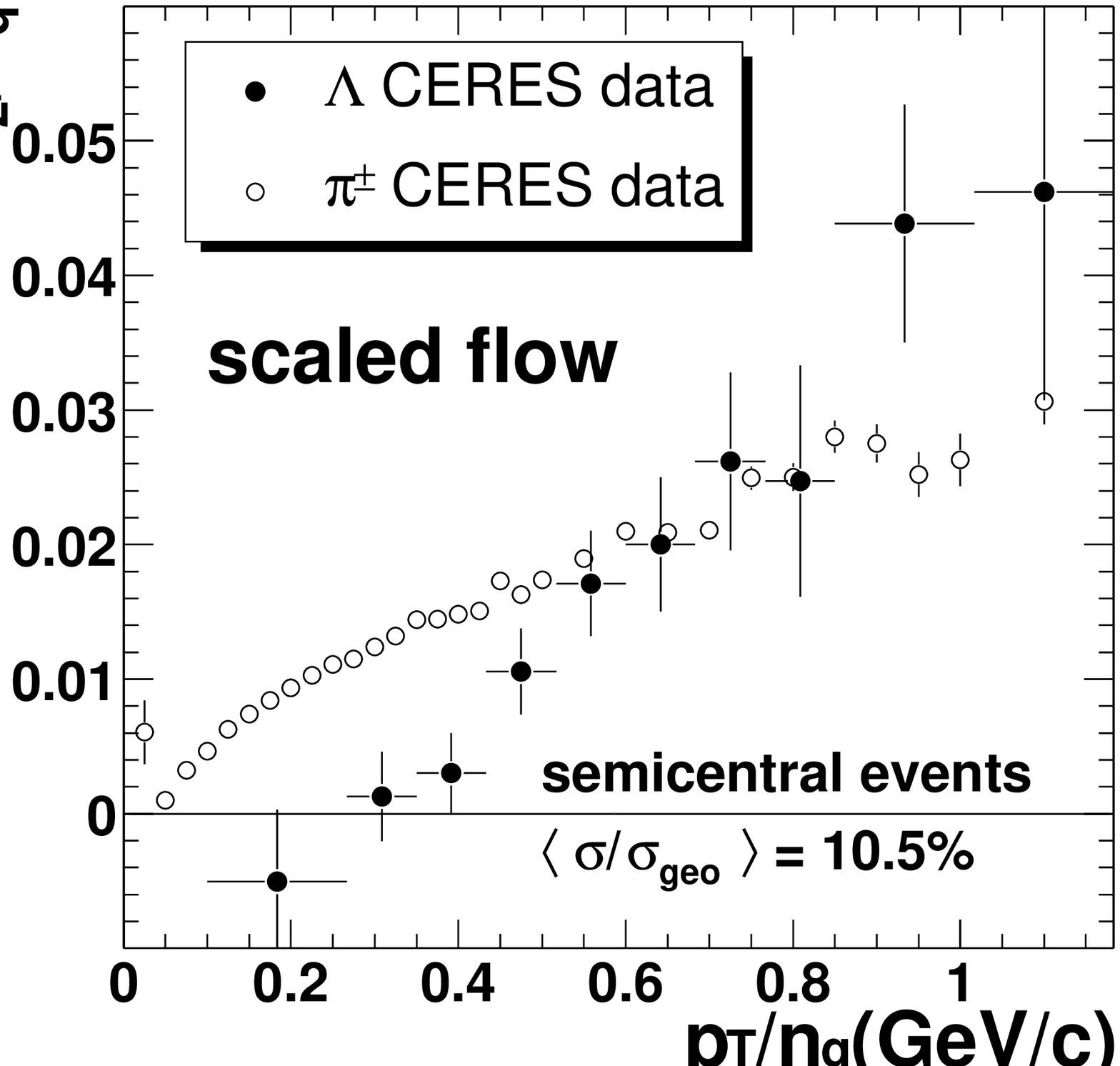}}
    \vspace*{-0.65cm}
    \caption{Comparison between $\Lambda$ and $\pi$ elliptic flow scaled to
    the number of constituent quarks.}
    \label{fig:scaled_quark_2}
  \end{minipage}
  \hspace{\fill}
  \begin{minipage}[c]{.48\textwidth}
    \centerline{\includegraphics[width=6.5cm]{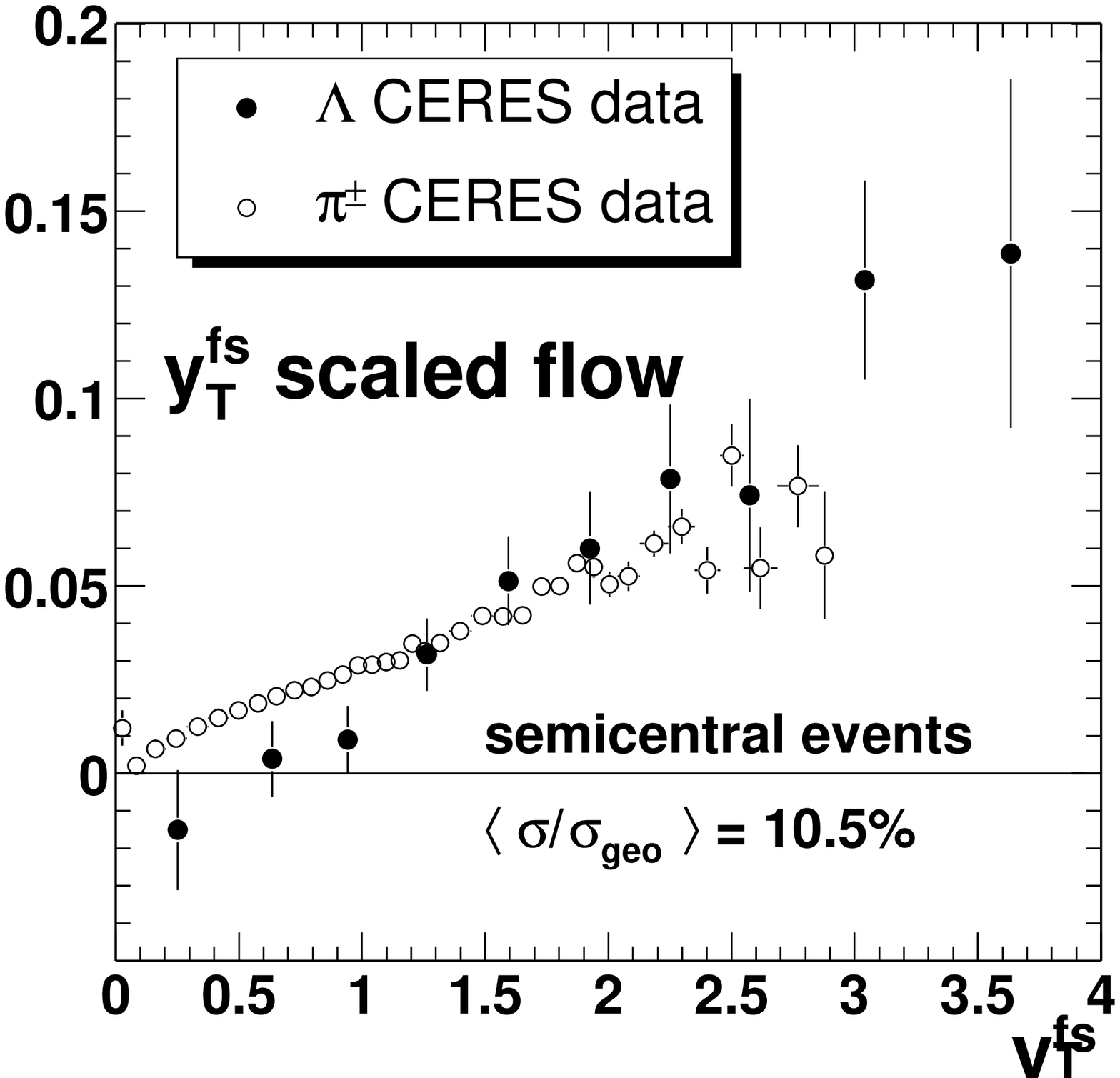}}
    \vspace*{-0.65cm}
    \caption{Comparison between $\Lambda$ and $\pi$ elliptic flow scaled with
    respect to $y_{T}^{fs}$ variable.}
    \label{fig:scaled_ytfs_QM}
  \end{minipage}
\end{figure}
\vspace*{-0.6cm}
$v_{2}(\pi) > v_{2}(\Lambda)$ , while it is opposite in case of high
$p_{T}$. Two flow scaling hypotheses were proposed. One by dividing $v_{2}$
and $p_{T}$ by the number of constituent quarks \cite{Pos} and the other one
using flavor transverse rapidity, defined with $y_{T}^{fs}=k_{m}y_{T}^{2}m$,
instead of the transverse momentum \cite{Tar}. The application of these
scaling hypotheses on CERES data is shown in Fig.~\ref{fig:scaled_quark_2} and
Fig.~\ref{fig:scaled_ytfs_QM}. The constituent quark scaling works
approximately for $p_{T} \ge 1.5$~GeV/c and fails for pions of lower
transverese momenta, as is observed at RHIC \cite{Pos}. The transverse
rapidity scaling is fullfilled reasonably well.

\end{document}